%% file: main.tex
\documentclass{vldb_workshop}
\pdfoutput=1
\usepackage{balance}  
\usepackage{times}
\usepackage{fullpage}
\usepackage{graphicx}
\usepackage{color}
\usepackage{subfigure}
\usepackage{multirow}

\usepackage{wrapfig}
\usepackage[noend]{algorithm}
\usepackage{algorithmic}
\usepackage{distribalgo}

\newcommand{\frappe}{\textsc{frapp\'{e}}}

\newenvironment{item*}%
 {\begin{itemize}%
   \setlength{\itemsep}{-5pt}%
   \setlength{\parsep}{-5pt}%
   \setlength{\topsep}{-5pt}}%
 {\end{itemize}}

\begin{document}

\title{FRAPP\'E: Fast Replication Platform for Elastic Services}


\numberofauthors{6}


\author{
\alignauthor
Vita Bortrnikov\\
       \affaddr{IBM Research}\\
       \email{vita@il.ibm.com}
\alignauthor
Gregory Chockler\\
       \affaddr{IBM Research}\\
       \email{chockler@il.ibm.com}
\alignauthor
Dmitri Perelman\\
        \affaddr{Technion, Israel Institute of Technology}\\
        \email{dima39@tx.technion.ac.il }
\and
\alignauthor
Alexey Roytman\\
        \affaddr{IBM Research}\\
        \email{roytman@il.ibm.com}
\alignauthor
Shlomit Shachor\\
        \affaddr{IBM Research}\\
        \email{shlomiti@il.ibm.com}
\alignauthor
Ilya Shnayderman\\
        \affaddr{IBM Research}\\
        \email{ilyashn@il.ibm.com}
}
\date{\today}

\maketitle

\newtheorem{thm}{Theorem}
\newtheorem{property}{Property}

\newcommand{\tup}[1]{%
        \relax\ifmmode
	           \langle #1 \rangle%
        \else
                $\langle$#1$\rangle$%
        \fi
}

\newcommand{\Config}{\ensuremath{\textit{Config}}}
\newcommand{\cnfgEst}{\ensuremath{\textit{cnfgEst}}}
\newcommand{\view}{\ensuremath{\textit{view}}}
\newcommand{\ballotNum}{\ensuremath{\textit{ballotNum}}}
\newcommand{\acceptedVal}{\ensuremath{\textit{acceptedVal}}}
\newcommand{\decidedVal}{\ensuremath{\textit{decidedVal}}}
\newcommand{\Proposal}{\ensuremath{\textit{Proposal}}}
\newcommand{\AcceptedValue}{\ensuremath{\textit{AcceptedValue}}}
\newcommand{\DecisionCandidate}{\ensuremath{\textit{DecisionCandidate}}}
\newcommand{\DecisionCandidates}{\ensuremath{\textit{DecisionCandidates}}}
\newcommand{\Set}{\ensuremath{\textit{Set}}}
 
\newfloat{interface}{htbp}{interface}
\floatname{interface}{Interface}

\newfloat{algo}{htbp}{algo}
\floatname{algo}{Algorithm}

\definecolor{Green}{rgb}{0.1,0.4,0.1}
\definecolor{Blue}{rgb}{0.1,0.1,0.6}
\definecolor{DarkBlue}{rgb}{0.2,0.2,0.5}
\definecolor{Cyan}{rgb}{0.5,0.2,0.5}
\newcommand{\comment}[1]{{\color{Green}{$\rhd$ #1}}}
\newcommand{\commentcolor}{\color{Blue}}
\newcommand{\typecolor}{\color{Cyan}}
\newcommand{\elcomment}[1]{\hfill{\comment{#1}}}
\newcommand{\funcname}[1]{\texttt{#1}}
\newcommand{\typename}[1]{\textbf{#1}}


\begin{abstract}
Elasticity is critical for today's cloud services, which must be able
to quickly adapt to dynamically changing load conditions and resource
availability. We introduce \frappe, a new consistent replication
platform aiming at improving elasticity of the replicated services
hosted in clouds or large data centers. In the core of \frappe\/ is a
novel replicated state machine protocol, which employs {\em
speculative executions}\/ to ensure continuous operation during the
reconfiguration periods as well as in situations where failures
prevent the agreement on the next stable configuration from being
reached in a timely fashion.  We present the \frappe's architecture
and describe the basic techniques underlying the implementation of our
speculative state machine protocol.
\end{abstract}

\thispagestyle{empty}

\input{intro}

\input{related}

\input{arch}


\input{impl}

\input{conc}


\bibliographystyle{abbrv}

\balance
\bibliography{references}



\end{document}

%% file: intro.tex
\section {Introduction}
\label{sec:intro}

Replicated state machine~\cite{lamport-state-machine} is an important
tool for maintaining integrity of distributed applications and
services in\\ failure-prone data center and cloud computing
environments. In massively multi-tenant settings of today's clouds,
large number of replica groups share the common hardware
infrastructure, and compete for limited resources. In order to be able
to dynamically adapt to changing resource availability, load
fluctuations, variable power consumption, and support better data
locality, the consistent replication layer must be {\em elastic},
i.e., being capable of supporting dynamically changing replication
groups with a minimum disruption to the service availability and
performance.

However, to preserve correctness across configuration\\ changes, the
reconfiguration protocol must ensure that the state machine execution
responsibilities have been transferred to the members of the new
configuration in an orderly fashion, and in particular, no new user
commands are executed in the new configuration before it has been
agreed by the members of the old one. The resulting throughput
degradation might be prohibitive if the rate of dynamic changes is
high. Furthermore, the service availability will suffer if the old
configuration is lost (e.g., due to a failure) before the agreement on
the new one has been completed.  For example, Amazon Web
Services~\cite{aws} must guarantee $99.99$'s availability, which
translates to less than $52$ minutes of unavailability a year. If a
significant portion of the service up-time the normal operation is
interrupted to execute the reconfiguration protocol, these
availability goals might not be met.

In this paper, we introduce a new replication platform, called
\frappe\/ ({\em Fast ReplicAtion Platform for Elastic services}),
which resolves the inefficiencies, and availability limitations
associated with dynamic reconfiguration. In the core of \frappe\/ is a
novel replicated state machine protocol, which employs {\em
speculative executions}\/ to ensure continuous operation during the
reconfiguration periods as well as in situations where failures
prevent the agreement on the next stable configuration from being
reached in a timely fashion.




Internally, our speculative state machine implementation is based on
the reconfigurable Paxos
approach~\cite{classic-paxos,paxos-made-simple,ReconfStateMachine}: i.e.,
configurations are treated as a part of the replicated state, and are
being agreed upon in their own consensus instances. However, in
contrast to reconfigurable Paxos, the command ordering can continue to
execute normally even if the agreement on the configuration relative
to which those commands will be ordered is still in progress. The key
observation is that the command ordering can be executed in an {\em
estimated}\/ configuration provided the validity of the speculative
decisions can be verified once the next {\em agreed}\/ configuration
becomes available.

To accomplish that, in \frappe, each replica maintains a {\em
branching}\/ command log (see Figure~\ref{fig:recon}) as opposed to a
linear log maintained by the standard replicated state machine
implementations. Whenever a replica learns of (or proposes) a new
speculative configuration, it creates a new log branch originating in
the log location associated with the agreement instance created for
that configuration. Each branch then proceeds to execute its own
independent sequence of the Paxos agreement instances as explained in
Section~\ref{sec:log}. Whenever a replica learns the outcome of the
agreement for a slot with one or more speculative branches, it stems
all branches originating at that slot except for the one whose
configuration is identical to the one that has been agreed (if
exists).


Speculatively agreed commands can be applied to the state either
immediately (at the risk of the possible future rollback), or when
their branch is validated against the agreed configuration. Since in a
common case, estimated configurations would coincide with those being
eventually decided, and the result of the reconfiguration agreement
will be available by the time the first speculative command is agreed
upon, reconfiguration will have a little impact on the overall command
throughput. Furthermore, since incoming commands can continue to be
ordered in an estimated configuration, the system availability will be
unaffected even when underlying failures prevent the configuration
agreement from completion. The global log consistency is guaranteed to
be restored once the real configuration is learnt, and the speculative
branches are verified against it.

Reconfiguration can be driven by a variety of high-level resource
management systems (such as placement controllers, load balancers, and
health-monitoring systems), or initiated through the administrative
inputs. In \frappe, the reconfiguration decisions are mediated through
the Configuration Manager component (see Figure~\ref{fig:sys-arch}),
which plans and triggers configuration changes, based on the current
configuration membership, and pluggable reconfiguration policies.

The rest of the paper is organized as follows. Related work is
discussed in Section~\ref{sec:relwork}. The \frappe\/ architecture is
described in Section~\ref{sec:arch}. Section~\ref{sec:impl} discusses
our speculative replicated state machine implementation, and
Section~\ref{sec:conc} concludes the paper.

%% file: related.tex
\section {Related work}
\label{sec:relwork}

Several approaches to alleviating reconfiguration bottleneck in
reconfigurable state machines have been proposed. The original idea by
Lamport, described
in~\cite{classic-paxos,paxos-made-simple,ReconfStateMachine}, and
implemented in SMART~\cite{Lorch_thesmart}, was to delay the effect of
the configuration agreed in a specific consensus instance by a fixed
number $\alpha$ of successive consensus instances. If the
configuration must take effect immediately, the remaining instances
can be skipped by passing a special ``window closure'' decree
consisting of $\alpha$ consecutive {\em noop}\/ instances. Although
this approach allows up to $\alpha$ consecutive commands to be
executed concurrently, choosing the right value of $\alpha$ is
nontrivial. On the one hand, choosing $\alpha$ to be too small may
under-utilize the available resources. On the other hand, large values
of $\alpha$ may not match the actual service reconfiguration rate
resulting in too frequent invocations of the window closure decrees
(which must complete synchronously).

Chubby~\cite{Burrows:2006:CLS:1298455.1298487} and
ZooKeeper~\cite{Hunt:2010:ZWC:1855840.1855851} expose
high-level synchronization primitives (respectively, locks and
watches) that can be used to implement a reconfigurable state machine
within the client groups. The solutions based on this approach are
however, vulnerable to timing failures, and therefore, either restrict
their failure model~\cite{chain-rep}, or rely on additional
synchronization protocols within the replication groups themselves to
maintain consistency~\cite{VerticalPaxos,spinnacker}.

Vertical Paxos~\cite{VerticalPaxos} removes the configuration
agreement overhead from the critical path by delegating it to an
auxiliary ``configuration master''. The reconfiguration involves an
extra step of synchronizing with the read quorums of all preceding
configurations causing throughput degradation. In addition, the
configuration master itself is implemented using the $\alpha$-based
reconfigurable Paxos protocol, and therefore, suffers from the
limitations similar to those discussed above.

Dynamic reconfiguration has been extensively studied in the context of
virtually synchronous group
communication~\cite{aleta,birman-joseph,esti,lynch}, and
reconfigurable read/write registers~\cite{Lynch02rambo:a,shraer}. The
reconfiguration protocols described in these papers do not aim to
support consistency semantics as strong as those of state machine
replication, and therefore do not directly apply in our context.
Birman et al.~\cite{VirtuallySynchronousMethodology} present a
replication framework unifying reconfigurable state machine and
virtual synchrony in which the normal operation is suspended during
the reconfiguration periods.

Optimistic and speculative approaches to mask the coordination latency
have been extensively studied in the past in a variety of contexts
(such as e.g., group communication~\cite{bankomat}, and database
replication~\cite{tutu}).  However, to the best of our knowledge,
speculative reconfigurable state machine replication studied in this
paper has not yet been addressed in the prior work.

%% file: arch.tex
\section{System Architecture}
\label{sec:arch}

\begin{figure}[t]
	\centering
	\includegraphics[width=0.45\textwidth]{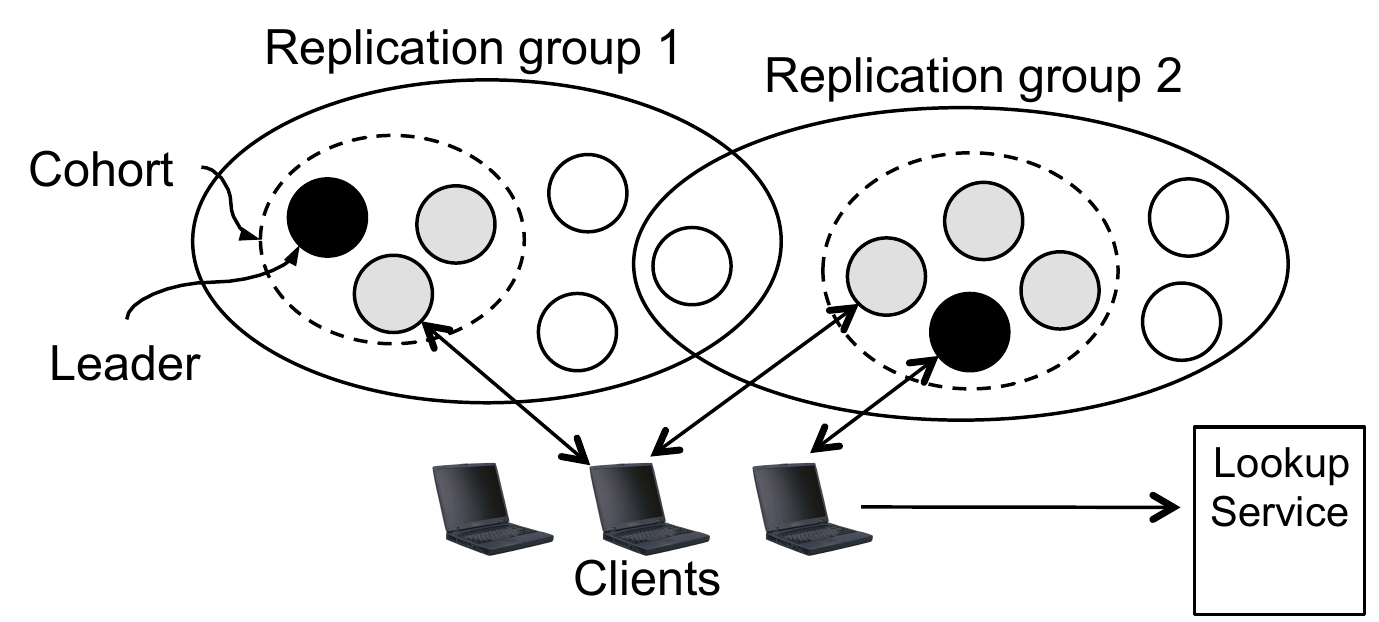}
	\caption{\small{The \frappe\/ Group Structure}}
	\label{fig:groups}
\end{figure}

The members of each elastic service cluster whose state is managed
through \frappe\/ are organized into {\em replication groups} (see
Figure~\ref{fig:groups}). Each member of the replication group can be
either {\em active}\/ or {\em idle}. The active members hold the
up-to-date copies of the service state, and are responsible for the
client and reconfiguration command ordering. The set of all currently
active members of a replication group form the group's {\em cohort}.
The ordering protocol is orchestrated by a distinguished member of the
cohort, called the {\em leader}. Although the idle processes do not
participate in the command ordering, they are nevertheless available
for serving the state transfer requests (see
Section~\ref{sec:state-transfer}). They can be taken off-line once it
is verified their copies of the service state have been propagated to
a sufficient number of the group members.

The current cohort configuration and leader identity within each
replication group is maintained by the cohort members, and propagated
to the idle members through a gossip-based protocol. This information
is made available to the clients, and newly joining replicas through
an external lookup service, such as DNS or LDAP, which maps each group
name to its cohort configuration and leader identity.

\begin{figure}[!ht]
	\centering
	\includegraphics[width=0.45\textwidth]{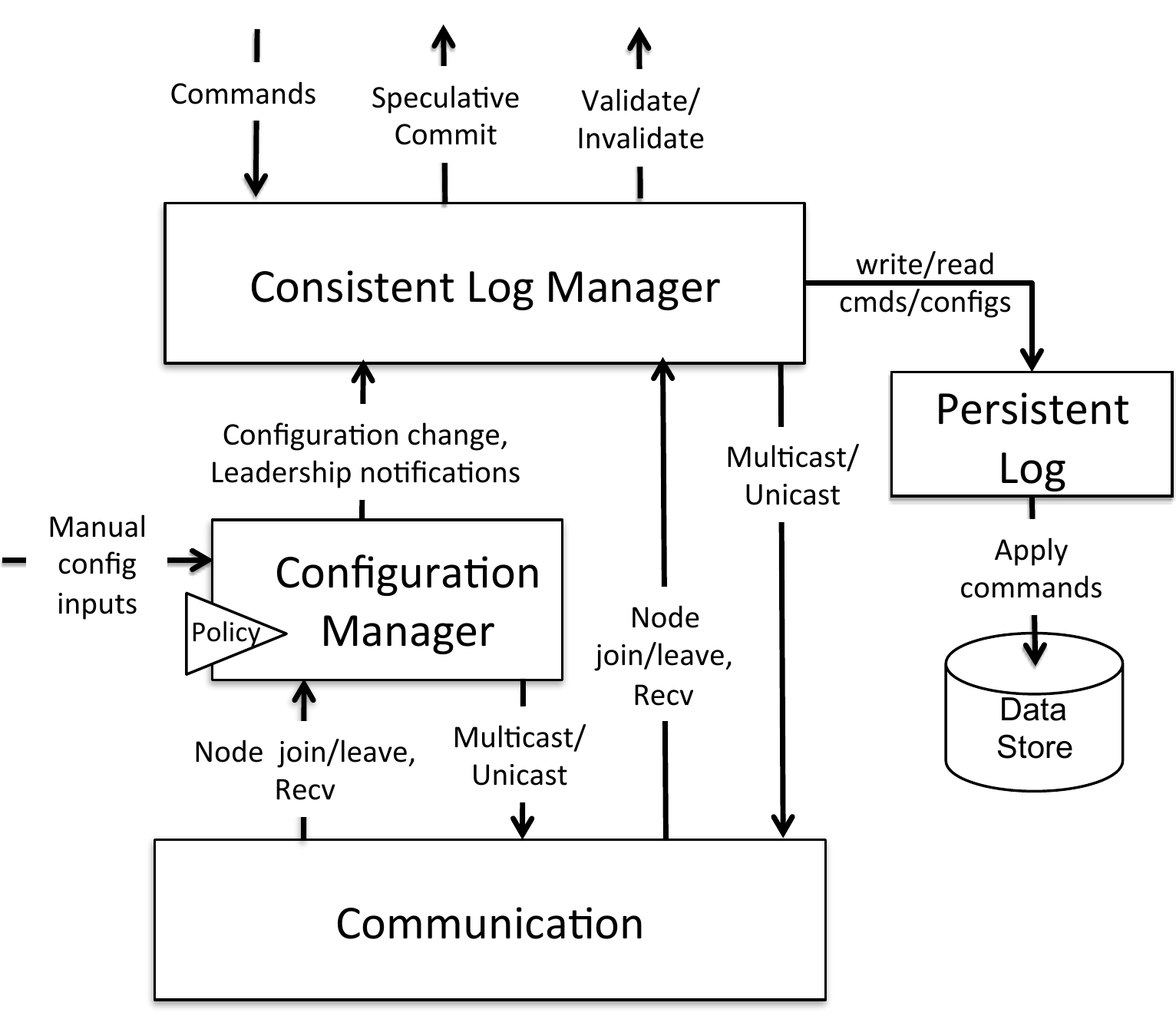}
	\caption{\small{The \frappe\/ Layers}}
	\label{fig:sys-arch}
\end{figure}

The layered structure of each individual \frappe\/ replica is depicted
in Figure~\ref{fig:sys-arch}.  The user commands are processed by the
Consistent Log Manager layer whose responsibility is to maintain
globally ordered command log with possible speculative branching as
explained in Section~\ref{sec:log}. The command log is persisted on
stable storage through the Persistent Log layer. Whenever the size of
a locally maintained log grows beyond a configured upper limit, a
portion of the globally ordered command prefix is clipped, and the
commands in this prefix are applied to the local copy of the
replicated state kept on the persistent data store.

Configuration Manager keeps track of the current cohort configuration,
and plans and triggers configuration changes, which are passed to
Consistent Log Manager for agreement. New cohort members are recruited
from the set of the idle members of the replication group. The
configuration change decisions are driven by the current view of the
replication group, cohort membership, and a pluggable policy.
Configuration Manager is also in charge of selecting the cohort
leader. Both Consistent Log and Configuration Managers utilize
services provided by the Communication layer for disseminating
protocol messages, and failure detection.


%% file: impl.tex
\section{Speculative State Machine}
\label{sec:impl}
In this section, we discuss the two main parts underlying our
speculative replicated state machine implementation, which are
speculative command ordering and state transfer protocol, in more
details.

\subsection{Command Ordering}
\label{sec:log}

\begin{figure*}[t!]
	\centering
	\includegraphics[width=0.7\textwidth]{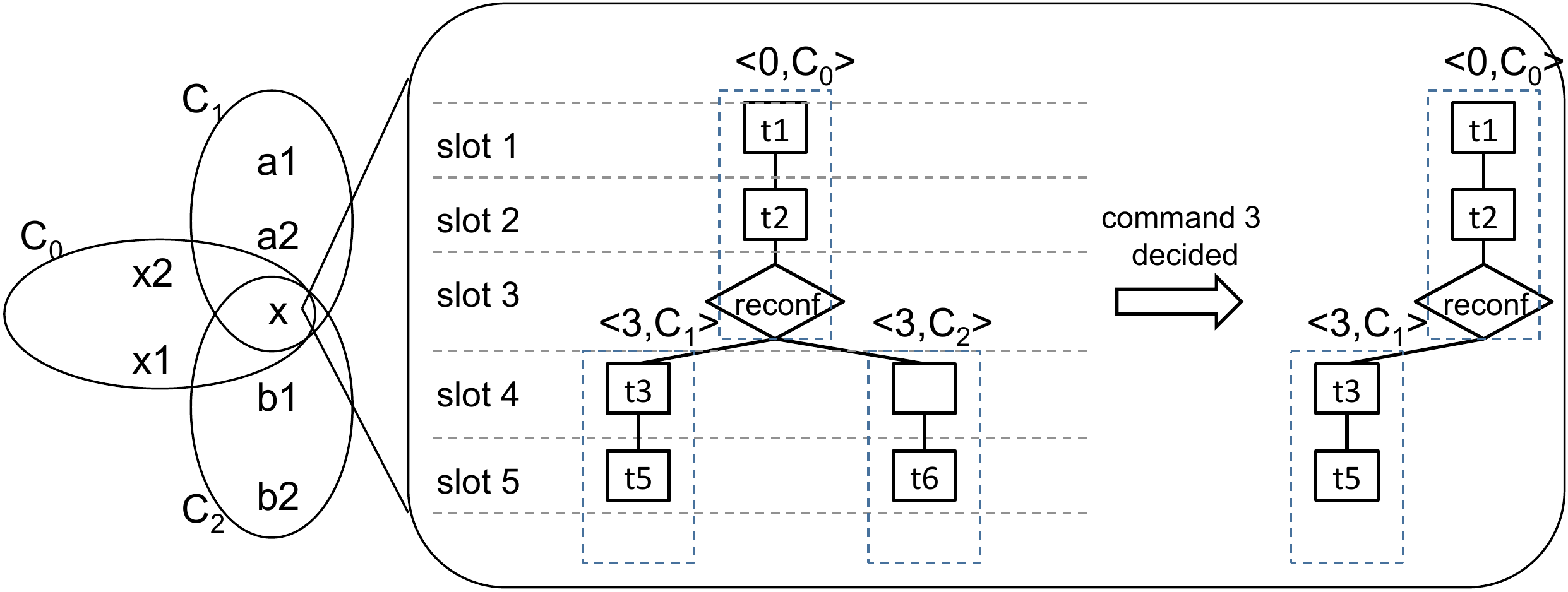}
	\caption{\small{Maintaining multiple branches during
        reconfiguration. Ballot and branch initiator components of the branch identifiers are omitted 
for clarity. Node $x$ participates in two speculative
        branches \tup{3,$C_1$} and \tup{3,$C_2$}. Once $x$ learns the
        configuration decided by command $3$, it prunes branch
        \tup{3,$C_2$} and merges branch \tup{3,$C_1$} into the
        trunk. }}
	\label{fig:recon}
\end{figure*}

Our command ordering implementation is based on the standard
reconfigurable Paxos
approach~\cite{classic-paxos,paxos-made-simple,ReconfStateMachine}
(also known as ``Horizontal Paxos''~\cite{VerticalPaxos}) augmented
with the speculative ordering support. As in standard Paxos, each
individual slot in the log is occupied by the user command chosen by
an instance of the agreement protocol. The agreement protocol is
orchestrated by a single process, called the {\em leader}, and
consists of two phases each one associated with a unique monotonically
increasing ballot number, and consisting of one round-trip message
exchange. The first phase chooses the value to propose in the slot,
and the second phase stores the chosen value at the
replicas. Agreement instances for each slot are independent of each
other, and can be executed in parallel.  The first phase can be shared
across multiple agreement instances provided they are mediated by the
same leader and use the same ballot number.  Thus, in the common
case when the leader remains live and connected for a long time,
each command is ordered within just two message delays, and multiple
command orderings are executed in parallel resulting in a high
throughput, low latency protocol.

Each command agreement is executed against a quorum of processes of
the current replica configuration. In Horizontal Paxos, a
configuration is treated as a part of the replicated service state:
i.e., the configuration for slot $i$ of the log is determined by the
latest configuration change command preceding $i$ in the log. Below,
we describe how this basic scheme is extended to allow {\em
speculative}\/ execution of the user commands relative to
configuration {\em estimates}\/ instead of agreed configurations. We
will focus on the novel aspects of the speculative state machine
implementation, and omit the details of Horizontal Paxos, which are
well-known.

\input{alg-pseudo-code}


The sequence of commands executed relative to a configuration estimate
$C$ is represented as a {\em branch}\/ in the global command log
starting at the position allocated for the agreement on $C$ (see
Figure~\ref{fig:recon}). Specifically, the branch spawned by a
reconfiguration command $cmd$ is uniquely identified by a triple
consisting of the following three components (lines
Algorithm~\ref{alg:ds-pseudo}.2--4): (1) slot number occupied by $cmd$
in the parent branch , (2) ballot number that was used to propose
$cmd$, and (3) identifier of the leader that proposed $cmd$. The
branch identifier is attached to all messages associated with the
Paxos agreement instances proposed within the branch, and is used to
route received messages to the correct instance of the Paxos protocol
(lines Algorithm~\ref{alg:replica-pseudo}.20--22).

Each branch executes a stream of the Paxos agreement instances, and
the agreement streams within different branches proceed independently
from each other (see Figure~\ref{fig:recon}). In particular, each
branch maintains its own set of the Paxos data structure, a subset of
which consisting of the maximum ballot number, the totally ordered
command prefix, and the next available agreement slot are exposed to
the speculative state machine
(lines Algorithm~\ref{alg:ds-pseudo}.9--11). All the live branches known to
a replica are kept in the set $branches$
(line Algorithm~\ref{alg:ds-pseudo}.13), and are linked together into a
tree-like structure through the $parent$ branch identifier associated
with each branch (line Algorithm~\ref{alg:ds-pseudo}.7).

A replica creates a new speculative branch and adds it to its local
$branches$ set (line Algorithm~\ref{alg:replica-pseudo}.32) whenever it
receives a \textsc{join} message carrying the new and parent branch
identifiers, and the new configuration that spawned the branch. The
replica then starts a new instance of the Paxos agreement for that
branch initializing it with the ballot number, and the next agreement
slot number extracted from the new branch identifier. This way, if the
parent and the new configuration share the same leader (which is a
common case in practice), the command agreements can proceed to
execute speculatively in the new branch under the same ballot, thus
maintaining the same command throughput across the two configurations.

A speculative branch $B$ is called {\em valid}\/ if the
reconfiguration command that spawned $B$ is agreed in its parent
branch. Conversely, $B$ is {\em invalid}\/ if the agreed command
occupying $B$'s parent branch slot at which $B$ is rooted is either
reconfiguration command for a different configuration, or a user
command. (This way, each branch in $branches$ can be either one of the
following three types: valid, invalid, or speculative.) The globally
ordered command prefix is represented by the state variable called
$trunk$ defined in line Algorithm~\ref{alg:replica-pseudo}.14. The
valid branches can be merged into $trunk$ provided they form a
continuous sequence rooted at a slot of $trunk$. This is accomplished
by the background task shown in Algorithm~\ref{alg:pb-pseudo}, which
traverses a series of valid branches (stored in the\\ $curBranch$
variable) switching to the next branch whenever the reconfiguration
command that spawned this branch becomes agreed and merged into the
trunk (lines Algorithm~\ref{alg:pb-pseudo}.42--46). At this point, all
other branches rooted at the same slot as well as the currently
traversed parent branch are recursively discarded (lines
Algorithm~\ref{alg:pb-pseudo}.44, \ref{alg:pb-pseudo}.47).

The speculative branch structure is exposed to the clients, and the
commands agreed in those branches with a special flag indicating that
their final status is not yet known. In addition, the clients are
notified whenever a new speculative branch is created. They can then
act upon this notification, e.g., by creating the snapshot of the
current state for the possible future rollback. The clients are
delivered a {\em validate}\/ notification for a command once it is
merged into $trunk$, and an {\em invalidate}\/ notification if the
branch to which the command belongs has been discarded.

\subsection{State Transfer}
\label{sec:state-transfer}

When the system transitions to a new configuration, the implementation
must ensure that the latest system state is transferred to the members
of the new configuration. Since the standard practice of interrupting
the user command execution for the duration of the state transfer
would cancel out the benefits of speculation, in \frappe, both state
transfer and speculative command ordering proceed in parallel as
explained below. 

An instance of the state transfer protocol is activated whenever a new
branch is created in response to a \textsc{join} message received by a
replica (line Algorithm~\ref{alg:replica-pseudo}.33).  Since the
instances of the state transfer protocol in all branches execute the
same logic, in the following, we limit our attention to the steps
taken by a single state transfer instance activated in response to
changing the cohort membership from $C$ to $C'$. Note that although
the replicas in $C \setminus C'$ might eventually leave the cohort
(i.e., become idle), they nevertheless remain available for serving
state transfer requests as long as they are on line.

If both $C$ and $C'$ share the same leader, then it can continue
ordering commands by executing only the second (``proposal'') phase of
the Paxos protocol using the same ballot number.  Otherwise, the new
leader must execute the first (``prepare'') phase of Paxos to
establish the new ballot number, and learn the latest global state
prior to resuming the normal operation.

In addition, each newly joined process $p\in C'\setminus C$ (if any)
determines which parts of $trunk$ of size $n$ are not reflected in its
local copy of the state, and initiates state transfer with the members
of $C$ to fill out the missing parts. If the replicas in $C$ are
unavailable or do not have up-to-date state (e.g., if state transfer
in the $C$'s branch is still in progress), $p$ will request the
missing information from the currently available members of the
replication group.  If $p$ looses some of the ordering decisions
reached in $C'$ prior to $p$'s join, it will request the missing
commands directly from the $C'$'s leader.














%% file: alg-pseudo-code.tex
\newcounter{alg:tree-pseudo:lines}
\begin{algo}[!ht]
\caption{Types and States for Replica $p_i$:} 
\label{alg:ds-pseudo}

\centering
\footnotesize
\begin{distribalgo}[1]

\INDENT {Record {\bf BranchID}:} \label{alg:lines:bid-begin}
	\STATE $\mathbb{N}$: $slotNum$ \elcomment {slot num of the reconfiguration command in the parent branch}
	\STATE Ballot: $bal$ \elcomment {initial ballot of the branch}
	\STATE PID: $branchInitiator$ \elcomment {identifier of the leader that proposed this branch's configuration}
\ENDINDENT

\medskip

\INDENT{Record {\bf PaxosBranch}:}
	\STATE BranchID: $bid$ \elcomment {identifier of this Paxos branch}
        \STATE BranchID: $parent$ \elcomment {identifier of the parent branch}
        \STATE PID[]: $config$ \elcomment {configuration of this Paxos branch}
	\STATE Ballot: $bal$ \elcomment {maximum ballot number locally known to this Paxos branch}
	\STATE Command[]: $cmdLog$ \elcomment {prefix of user and reconfiguration 
commands ordered by this Paxos branch}  
        \STATE $\mathbb{N}$: nextSlot \elcomment {next available command slot locally known to this Paxos branch}
\ENDINDENT

\medskip

\INDENT {{\bf Replica State Variables}:}
  \STATE Set of PaxosBranch: $branches$ \elcomment {locally known live Paxos branches}\\
  \STATE Command[]: $trunk$ \elcomment {global total order prefix}\\
\ENDINDENT

\medskip

\INDENT{{\bf Replica State Initialization}:}
  \STATE Let $B_0=\tup{b_0, \bot, C_0, bal_0, \tup{}, 1}$ 
where $C_0$ is the initial configuration,
$b_0=\tup{0, bal_0, q}$, and $q$ is a deterministically chosen member of $C_0$.
  \STATE $branches=\{B_0\}$
  \STATE $trunk=\tup{\tup{\textsc{recon}, \bot, b_0, C_0}}$
  \STATE Start an instance of Paxos protocol for branch $B_0$ 
\ENDINDENT
\setcounter{alg:tree-pseudo:lines}{\value{ALC@line}} 
\end{distribalgo}
\end{algo}

\begin{algo}[!ht]
\caption{Message Handlers for Replica $p_i$:} 
\label{alg:replica-pseudo}

\centering
\footnotesize
\begin{distribalgo}[1]
\setcounter{ALC@line}{\value{alg:tree-pseudo:lines}}

\INDENT {{\bf Upon receiving} $m=\tup{\textsc{user}, bid, \cdot}$ or $\tup{\textsc{recon}, bid, \cdot}$:} \label{alg:line:msgrcv-begin}
	\INDENT {{\bf if} ($\exists B\in branches$, $B.bid=bid$) {\bf then}} 
                \STATE Pass $m$ to the Paxos instance for branch $B$ \label{alg:line:msgrcv-end}
        \ENDINDENT
\ENDINDENT

\medskip

\INDENT {{\bf New user command } $cmd$:}\label{alg:line:requests}
	\INDENT {For each branch $B \in branches$ where $p_i$\\ is the proposer for $x$:}
                \STATE Propose $\tup{\textsc{user}, B.bid, cmd}$ in the instance of Paxos protocol 
for $B$         
        \ENDINDENT
\ENDINDENT

\medskip

\INDENT {{\bf New reconfiguration command} $cmd=\tup{C}$:}
	\INDENT {For each branch $B \in branches$ where $p_i$\\ is the proposer for $B$:}
                \STATE Let $x=\tup{B.nextSlot, B.bal, p_i}$
                \STATE Send $\tup{\textsc{join}, B.bid, x, C}$ to all members of $C$                
                \STATE Propose $\tup{\textsc{recon}, B.bid, x, C}$ in the instance of Paxos protocol
for branch $B$
        \ENDINDENT
\ENDINDENT

\medskip

\INDENT {{\bf Upon receiving}: $\tup{\textsc{join}, bid, x, C}$:}
	\STATE Add branch $X=\tup{x, bid, C, x.bal, \tup{}, x.slotNum+1}$\\ to $branches$
        \STATE Start new instance of Paxos protocol for branch $X$, and initiate state transfer
(see Section~\ref{sec:state-transfer})

\ENDINDENT

\setcounter{alg:tree-pseudo:lines}{\value{ALC@line}} 
\end{distribalgo}
\end{algo}

\begin{algo}[!ht]
\caption{Total Order Construction at Replica $p_i$:} 
\label{alg:pb-pseudo}

\centering
\footnotesize
\begin{distribalgo}[1]
\setcounter{ALC@line}{\value{alg:tree-pseudo:lines}}

\INDENT {{\bf Task} Construct Total Order:}
        \STATE $curBranch := B_0$
        \STATE $next := 1$
        \INDENT {{\bf while} (true) {\bf do}:}
                \STATE $idx := next - curBranch.bid.slotNumber$
                \STATE Block until $idx \leq \textsf{length}(curBranch.cmdLog)$
                \INDENT {{\bf while} ($idx \leq \textsf{length}(curBranch.cmdLog)$) {\bf do}:}
                        \STATE $cmd := curBranch.cmdLog[idx-1]$
                        \STATE $B := \{b \in branches : b.parent=curBranch \wedge b.slotNumber=next\}$
                        \INDENT {{\bf if} ($cmd=\tup{\textsc{recon}, \cdot, x, \cdot}\ \wedge$\\
                                \ \ \ \ \ \ \ \ \ \ $\exists b\in B$ : $b.bid=x$) {\bf then}}
                                \STATE Discard $curBranch$, terminate Paxos instance associated with $curBranch$
                                \STATE $curBranch := b$
                                \STATE $B := B \setminus \{b\}$
                        \ENDINDENT
                        \STATE Recursively discard all branches in $B$
                        \STATE Append $cmd$ to $trunk$
                        \STATE $next := next + 1$
                \ENDINDENT
        \ENDINDENT
\ENDINDENT
\setcounter{alg:tree-pseudo:lines}{\value{ALC@line}} 
\end{distribalgo}
\end{algo}

%% file: conc.tex
\section{Conclusions}
\label{sec:conc}

In this paper, we have introduced \frappe\/ -- a replication platform
designed for supporting elastic services and applications hosted in
clouds and large data centers. The key to the \frappe's efficiency is
the new replicated state machine protocol, which is capable of
avoiding the delays associated with the replication group
reconfiguration by ordering commands speculatively in parallel to
reaching agreement on the next stable configuration. We have discussed
the \frappe's architecture, API, and core ideas underlying our
speculative state machine implementation. 

\frappe\/ is being developed at IBM Research, and slated to be
included into the future platform-as-a-service offerings as a
foundational tool. The experimental performance study of the initial
prototype are currently under way, and the results will be available
in the full version of the paper.